# Highly sensitive macro-scale diamond magnetometer operated by dynamical decoupling sequence with coplanar waveguide resonator

Y. Masuyama[1], K. Mizuno[1], H. Ozawa[1], H. Ishiwata[1], Y. Hatano[2], T. Ohshima[3], T. Iwasaki[1], M. Hatano[1]

[1] *Department of Electrical and Electronic Engineering, Tokyo Institute of Technology, Meguro, Tokyo, Japan*

[2] *Institute for Protein Research, Osaka University, Osaka, Japan*

[3] National Institutes for Quantum and Radiological Science and Technology, Takasaki, Gunma, Japan

Ultimate sensitivity for quantum magnetometry using nitrogen-vacancy (NV) centers in diamond is limited by number of NV centers and coherence time. Microwave irradiation with a high and homogeneous power density for a large detection volume is necessary to achieve highly sensitive magnetometer. Here, we demonstrate a microwave resonator to enhance the power density of the microwave field and an optical system with a detection volume of $1.4\times10^{-3}$ mm$^3$. The strong microwave field enables us to achieve 48 ns Rabi oscillation which is sufficiently faster than the phase relaxation time of NV centers. This system combined with a decoupling pulse sequence, XY16, extends the spin coherence time ($T_2$) up to 27 times longer than that with a spin echo method. Consequently, we obtained an AC magnetic field sensitivity of 10.8 pt/$\sqrt{\mathrm{Hz}}$ using the dynamical decoupling pulse sequence.

Ensemble nitrogen-vacancy (NV) centers in diamond are expected to reach a magnetic sensitivity in the femto-tesla range at room temperature, required for biological/medical applications including magnetoencephalography [1-4]. The sensitivity of diamond magnetometer is determined by number of the NV centers, spin coherence time, and detection efficiency [5, 6]. Pioneering work reported by Wolf et al. achieved the highest AC magnetic field sensitivity of 900 fT/$\sqrt{\mathrm{Hz}}$ by enlarging the detection volume to increase the number of the NV centers used for detection. Wolf *et al.* achieved a detection volume of $8.5\times 10^{-4}$ mm$^3$, a high detection efficiency of 65 %, and a spin coherence time of 100μs obtained by a spin echo sequence [7]. Dynamical decoupling of the large detection volume for extension of the spin coherence time beyond spin echo limit is necessary to extend its sensitivity limit. As for a single NV center, the dynamical decoupling sequences have been utilized to improve the spin coherence time and thus a magnetic sensitivity [8]. For driving the large volume of the NV ensembles, spatial inhomogeneity of the microwave field creates different Rabi oscillations of each NV center, causing rapid decays of the average signal from the NV ensemble. Therefore, the spatial homogeneity of the microwave field in the decoupling sequence becomes important. Furthermore, the Rabi oscillation faster than phase coherence time $T_2$* is critical to reduce the control error of the dynamical decoupling pulses. Thus, uniform and strong microwave driving is essential to achieve the dynamical decoupling for the large detection volume. In previous reports, the microwave irradiation of the large detection volume of the ensemble NV centers was performed by means of a spin echo sequence only [7, 9].

In this study, we constructed an efficient microwave and optical system and applied dynamical decoupling pulse sequences to an ensemble of NV centers in a large detection volume of $1.4\times10^{-3}$ mm$^3$. We used a coplanar waveguide resonator (CWR) with a wide center electrode for strong and uniform microwave irradiation. CWR enhances the microwave field only at the targeted frequency bandwidth and reflects the remaining wideband noise generated by the microwave amplifier. The strong microwave field allowed us to apply a robust dynamical decoupling pulse sequence called XY16 [10] which extended the coherence time of the ensemble NV centers up to 240 μs, which was 27 times longer than that obtained by the spin echo sequence. Consequently, we obtained an AC magnetic field sensitivity of 10.8 pT/$\sqrt{\mathrm{Hz}}$.

The microwave and optical system used in our experiment is shown in Fig. 1(a). The diamond sample containing NV centers was placed between the microwave resonator and a compound parabolic concentrator (CPC, Edmund Optics) [7]. A copper line on an FR4 substrate works as CWR [Fig.1 (b)]. A permanent magnet to fix the quantization axis and an AC field coil to evaluate the magnetic sensitivity were placed behind the resonator. The fluorescence from the NV centers was collected through the CPC and detected using a photodiode (referred to PD1, Thorlab FDS100). This configuration enabled us to apply the microwave field perpendicular to the NV axis along the [111] direction without interfering the laser beam passing from a substrate side. A Ib-type diamond substrate (2×2× 0.3 $mm^3$) was exposed to electron beam irradiation with a dose of $3.7\times10^{17}$ $cm^{-2}$and annealed, leading to an NV density of approximately $5 \times 10^{18}$ $cm^{-3}$. We controlled a TTL switch (Mini-Circuits ZASWA-2-50DR+) and an acousto-optic modulator (Gooch & Housego # 35250-.2-.53-XQ) using a pulse generator (Tektronix DTG5274) to generate the pulse sequence of the microwave and laser. After the microwave pulse sequence, a 400 μs measurement pulse of 532 nm laser was irradiated to the NV centers in diamond. The polarization of the laser at 700 mW power level was optimized to most efficiently excite the (111) component of the NV centers. The laser beam was split into two passes. One was aligned to the diamond sample for the measurements, while another pass was guided to a photodiode (referred to PD2, Thorlab FDS100) to compensate the fluctuation of the laser power. The laser beam to the sample was focused with a beam diameter of about 30 μm. The detection volume of the sample was ~ $1.4\times10^{-3}$ $mm^3$ with this configuration. The fluorescence signal from the NVs was amplified by a home-built amplifier after subtracting the signal of PD2 from that of PD1. The amplified signal was recorded by an 8-bit resolution oscilloscope (LeCroy Waverunner 640Zi) and analyzed to distinguish the spin state of the NV centers.

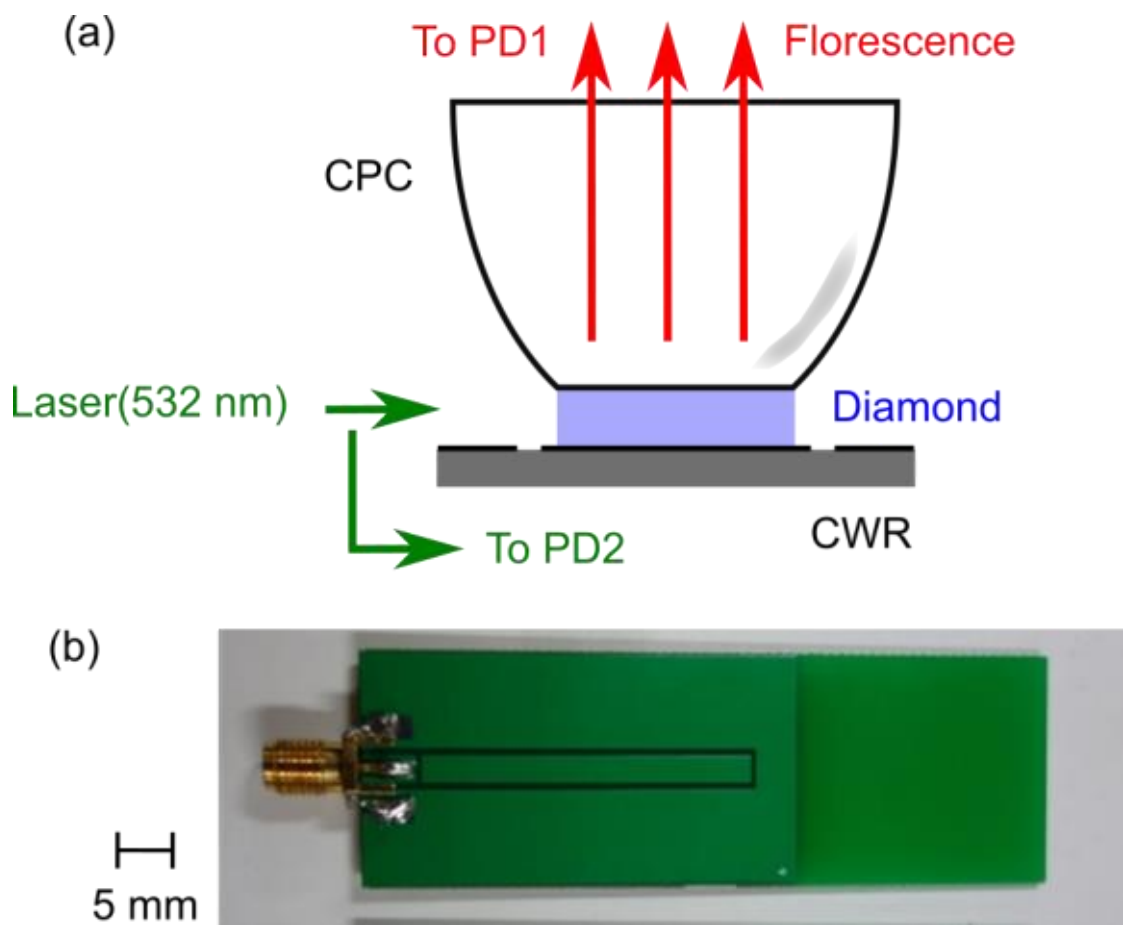

FIG. 1. The microwave and optical system of the diamond quantum magnetometer with large detection volume of nitrogen-vacancy centers. (a) Schematic of the setup of the diamond containing NV centers (~$5 \times 10^{18}$ $cm^{-3}$), photodiode (PD), compound parabolic concentrator (CPC), coplanar waveguide resonator (CWR). (b) Photograph of the CWR. The diamond sample was attached at the center of the central line.

We used the lowest mode of the resonator with a resonant frequency of 2.832 GHz to operate the NV centers. The quality factor without the diamond sample was 27 which corresponds to a cavity linewidth of 104 MHz. The resonator can effectively drive the (111)-oriented NV centers. Figures 2(a) and (b) show simulation results of the resonant magnetic field around the center of the resonator calculated by a finite element method (COMSOL 5.3). Figure 2(a) shows a two-dimensional plot of the magnetic field of CWR. We can clearly see that the magnetic field concentrates on the center of the central electrode. Compared with a ring-type microwave resonator [11] which was modified for a large detection volume and a 20 μm thin wire which was often used for a confocal microscope setup [12], our resonator allows maximum microwave field strength for driving a large volume [Fig. 2(b)].

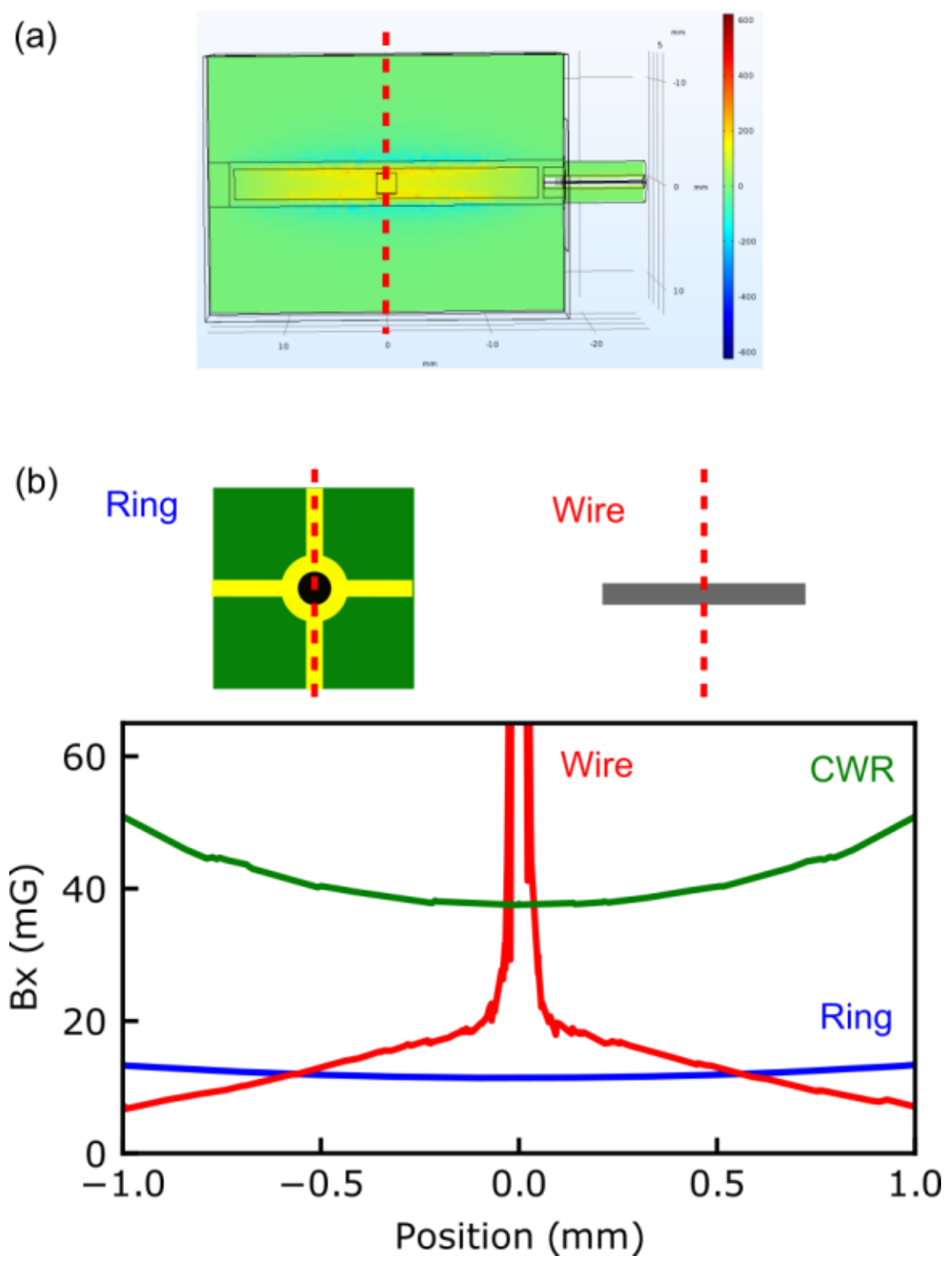

FIG. 2. Simulation of microwave resonators. (a) Two-dimensional plot of the magnetic field of the CWR. (b) Magnetic field amplitude of CWR (green), ring-type (blue), and wire (red). Each line shows the amplitude crossing the strongest amplitude point on the resonance.

We identified the resonance frequency of the (111) component of the NV centers by continuous wave optically detected magnetic resonance (CW-ODMR) [Fig.3 (a)]. A magnetic field of 2 mT was applied parallel to the NV axis by the permanent magnet. The resonant frequency of the (111) component at 2.8088 GHz was used for the following time-domain experiments. First, we measured Rabi oscillation of the ensemble NV centers. Figure 3(b) shows Rabi oscillation of the ensemble NV centers. Measured state flip (π pulse) time was 48 ns which was much faster than the phase coherence time $T_2^*$ of 150 ns. Note that the current Rabi frequency is limited by the saturation of the microwave amplifier.

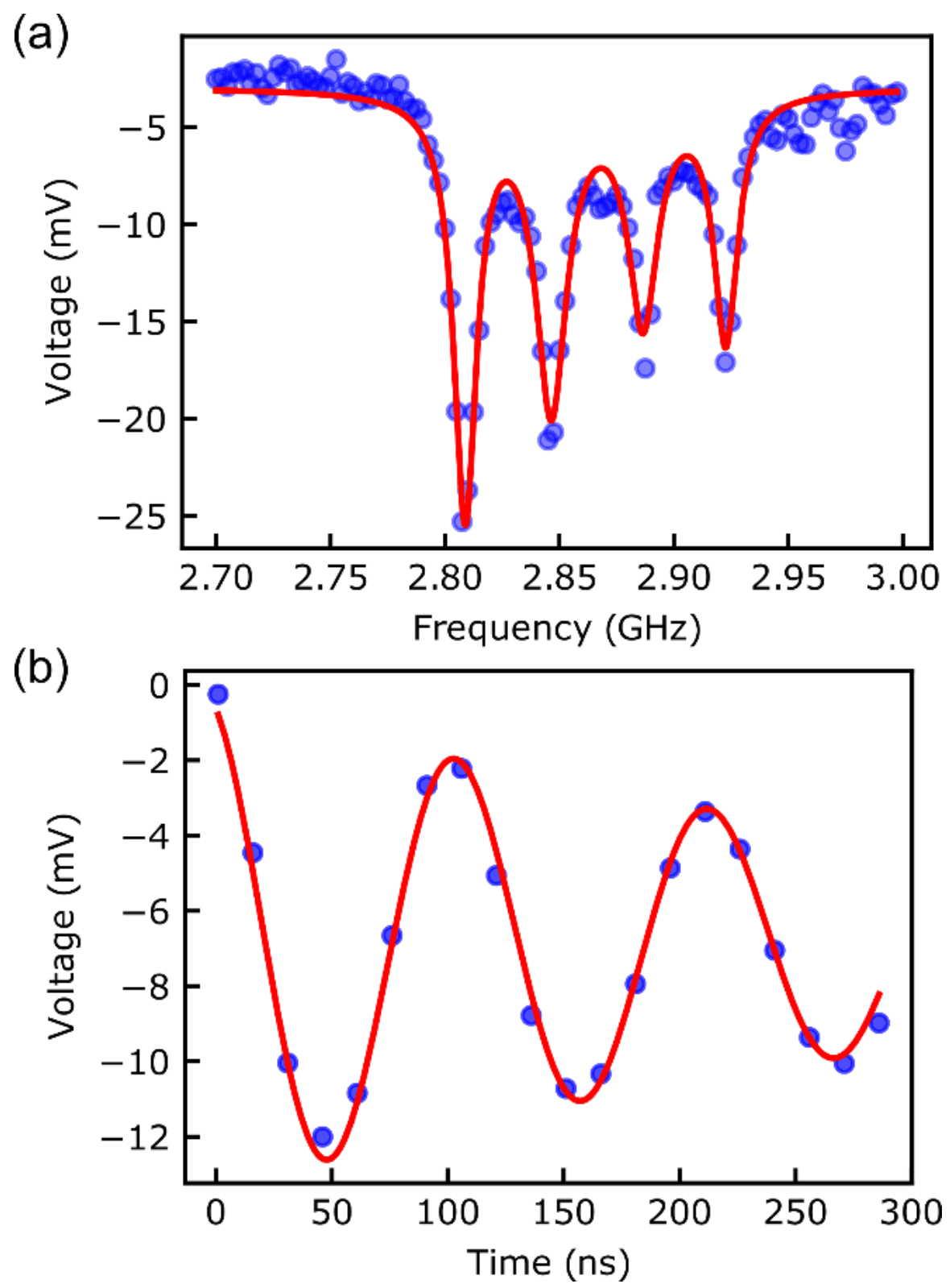

FIG. 3. (a) CW-ODMR spectrum (dots) and Lorentzian fit (solid line). The resonance dips were split by a 2 mT magnetic field. (b) Rabi oscillation (dots) and damped sine fits (solid line).

This fast π pulse for the ensemble NV centers in large detection volume allowed us to apply the XY16-N pulse sequence which is a N set of 16 rectangular π pulses of 47 dBm consisting of x-y-x-y-y-x-y-x-x-y-x-y-y-x-y-x where x and y indicate the rotation axis of the π pulse and x and y indicate the inverse rotation of x and y [Fig.4 (a)], respectively [10]. The XY16 pulse sequence is robust against amplitude error [13]. Similar to the previous report [7], two pulse sequences were implemented for the noise reduction, as shown in Fig. 4(a). In this study, however, the last π/2 pulse in each sequence was applied as (π/2) and (-π/2) for the first and second sequence, respectively. The data analysis was performed as follows.

$$S = (S1 - R1) - (S2 - R2), \qquad (1)$$

where the width of S1 and S2 (signal regions possessing the information on the spin state of the NV center) was 10 μs, the width of R1 and R2 (reference regions to calibrate the fluctuation of the laser power) was 50 μs [Fig. 4 (a)]. The subtraction R from S and (S1- R1) – (S2 – R2) are used to calibrate the fluctuation of the laser power and the microwave power, respectively. This calculation enables us to achieve scaling of the magnetic sensitivity with a square root of the measurement time $T_{seq}$ [7].

Figure 4(b) shows the normalized signals for the pulse sequences of the spin echo method and the XY16-N. The phase coherence time $T_{2,echo}$ of the NV centers measured by the spin echo method was 9 μs. The coherence time was apparently extended upon the application of the XY16-N sequences and reached 240 μs (N=16). Figure 4(c) shows the $T_2$ enhancement normalized with $T_{2,echo}$. The combination of the microwave resonator and the XY16 pulse sequence improved the phase coherence time of the ensemble NV centers by a factor of 27 compared to the spin echo method. We achieved application of the dynamical decoupling pulse sequence beyond spin echo method to the large detection volume of the NV centers for the first time. The experimental data are below the theoretical limit of the extension which is proportional to $N_{pls}{}^{2/3}$, where $N_{pls}$ is the total pulse number [14] . This difference was presumably caused by the control error of the NV center state and/or miss-alignment between the NV axis and DC magnetic field by the permanent magnet. Thus, an even longer spin coherence time can be expected by further tuning the experimental conditions.

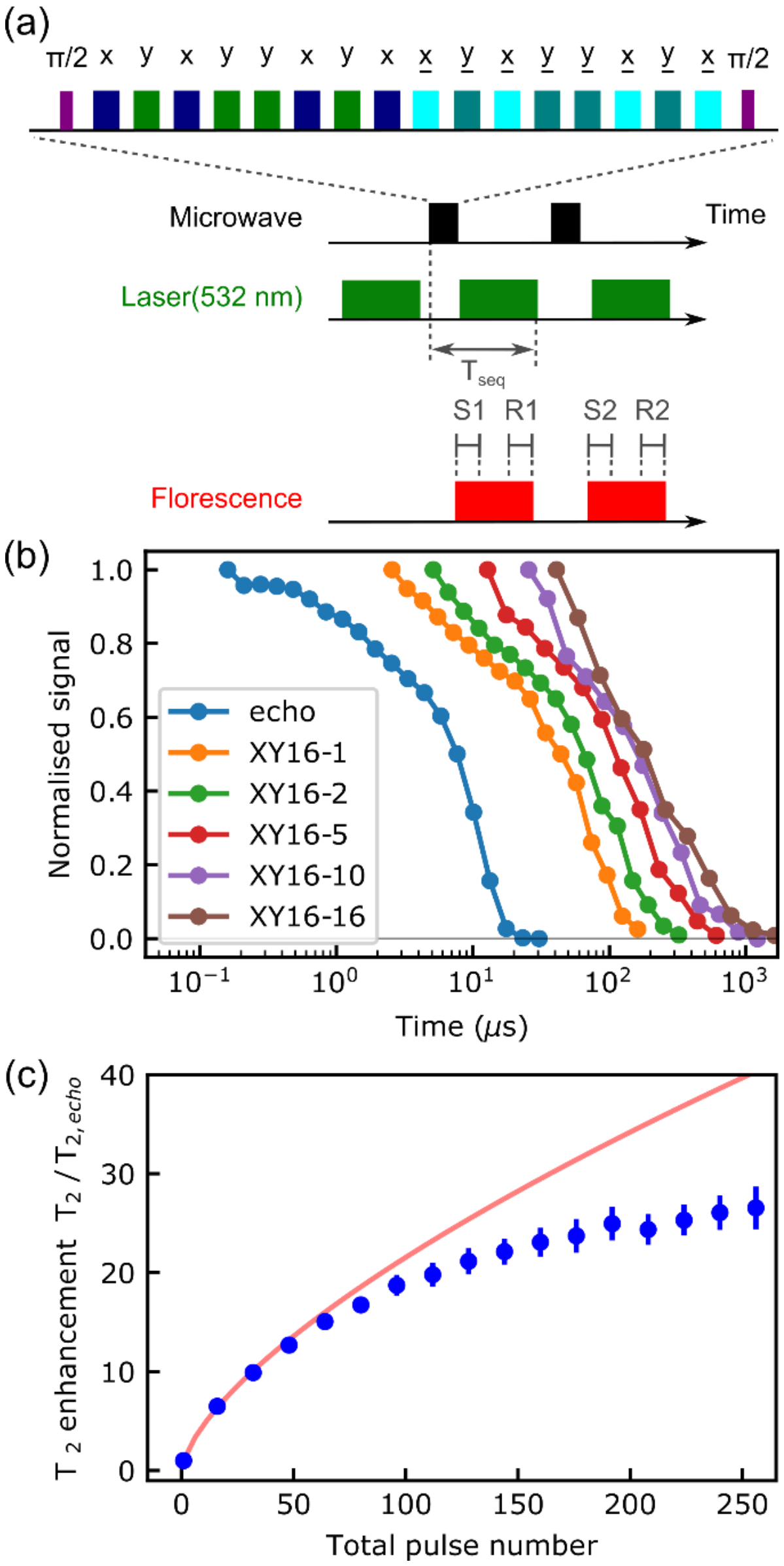

FIG. 4. Phase coherence time using dynamical decoupling pulse sequences. (a) Pulse sequence for XY16-1. S1, R1, S2, and R2 denote the regions for the signal processing in eq.(1). The last pi/2 pulse of the second set is pi/2. (b) Normalized decay signals using spin echo method, XY16-1, XY16-2, XY16-5, XY16-10, and XY16-16 sequence. (c) Enhancement factor of the coherence time $T_2$ / $T_{2,echo}$ as a function of the total pulse number $N_{pls}$. The solid line shows the theoretical limit proportional to $N_{pls}{}^{2/3}$ [14]. The spin coherence time of 9 µs by spin echo method ($T_{2,echo}$) was extended to 240 µs by the XY16-16 sequence.

Finally, we evaluated the sensitivity of the AC magnetic field [8] using the XY16-15 pulse sequence. We repeated the XY16-15 pulse sequence with different amplitudes of the AC magnetic field [Fig.5 (a)]. The frequency of the applied AC magnetic field was set to 362 kHz. The maximum sensitivity of the signal response, max |dS/dB| = 320000 A/V, was obtained close to the origin of the plot. The standard deviation of the signal δS for the single measurement was 89.4 μV. Thus, the magnetic sensitivity of the AC magnetic field was

$$\eta = \frac{\delta S}{\max|\partial S/\partial B|}\sqrt{T_{seq}} = 10.8\ \mathrm{pT}/\sqrt{\mathrm{Hz}}\ , \qquad (2)$$

where $T_{seq}$ is 1.47 ms. Figure 5 (b) shows the resolution of the magnetic field $\frac{\delta S}{\max|\partial S/\partial B|}$ as a function of the time. The minimum magnetic field reached 2.7 pT for 74 seconds corresponding to the 50000 times averaging. In the current setup, the magnetic sensitivity is limited by the standard deviation of the signal, which will be improved by utilizing a high-performance detector. In addition, from the viewpoint of the sensor materials, perfectly aligned NV ensembles along the [111] direction [15] will also improve the magnetic sensitivity with a higher ODMR contrast.

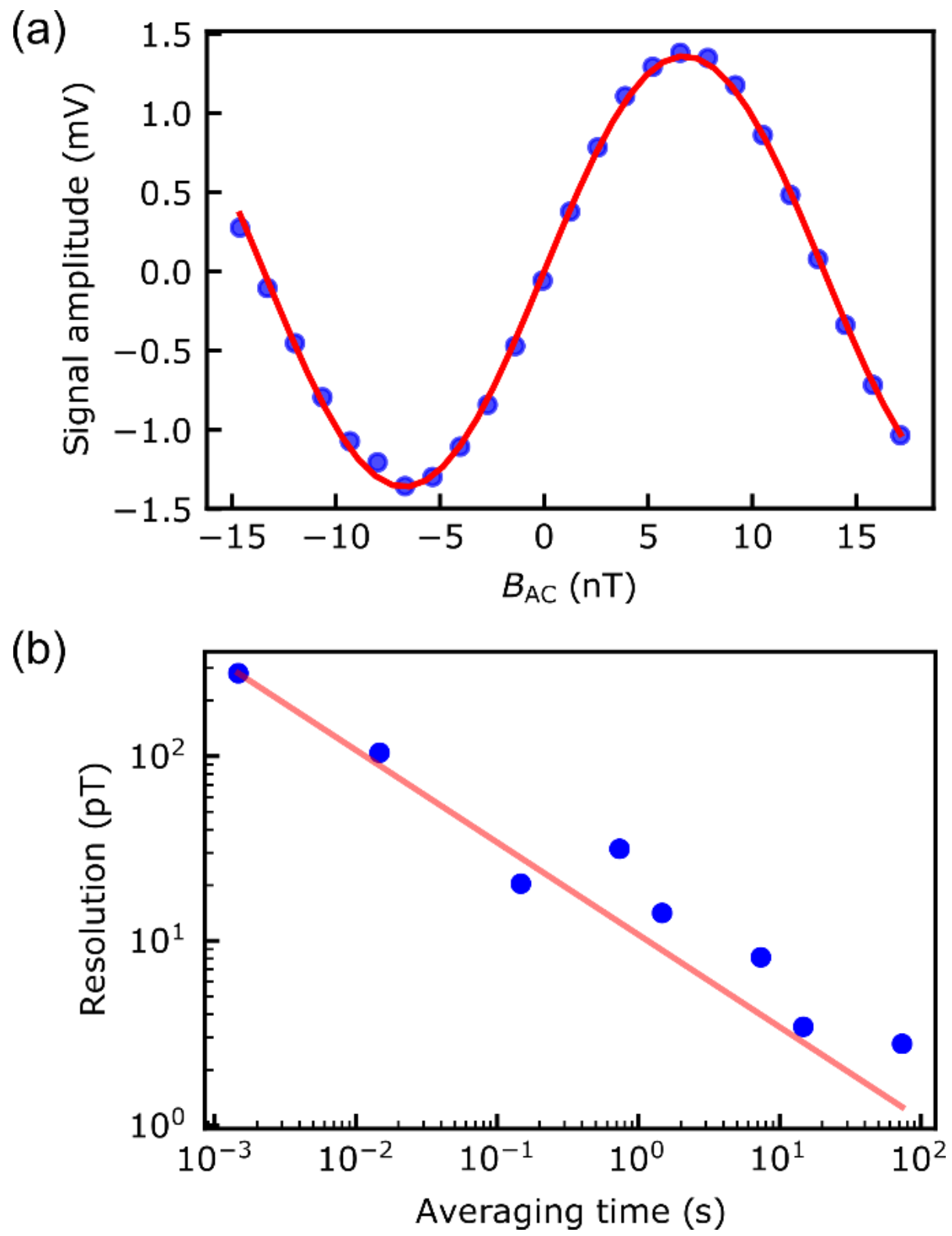

FIG. 5. AC magnetic sensitivity using dynamical decoupling sequence. (a) Dependence of the signal on the AC magnetic field amplitude. The solid line is a sine fit. (b) Magnetic field resolution as a function of the averaging time. The solid line shows the scaling of $T_{seq}^{-1/2}$.

In conclusion, we developed a microwave resonator to drive the ensemble NV centers in a large detection volume ($1.4\times10^{-3}$ $mm^3$). The spin coherence time with a dynamical decoupling sequence of XY16-16 reached 240 μs, which is 27 times longer than that obtained by the spin echo method. We obtained an AC magnetic field sensitivity of 10.8 pT/$\sqrt{\mathrm{Hz}}$ using the XY16-15 sequence. Further improvement of the magnetic sensitivity is expected by improving the signal stability and the signal contrast.

The authors would like to acknowledge Professor J. Wrachtrup for fruitful discussions. This work was supported by JST CREST Grant Number JPMJCR1333 and JSPS KAKENHI Grant (No. JP17H01262).